\documentclass[prc,ams,twocolumn,showpacs,superscriptaddress]{revtex4} 
\usepackage{graphicx}
\usepackage[dvips]{color}
\usepackage{colordvi}
\newcommand{\beq}{\begin{equation}}
\newcommand{\eeq}{\end{equation}}

\newcommand{\be}{\begin{eqnarray}}
\newcommand{\ee}{\end{eqnarray}}
\newcommand{\bes}{\begin{eqnarray*}}
\newcommand{\ees}{\end{eqnarray*}}
\newcommand{\et}{{\em et al.}}

\newcommand{\ceep}{$^{12}$C(e,e'p)}
\newcommand{\heep}{H(e,e'p)}
\newcommand{\emiss}{$E_{\rm m}$}
\newcommand{\pmiss}{$p_{\rm m}$}
\def\lsim{\mathrel{\rlap{\lower2.5pt\hbox{\hskip1pt$\sim$}}
    \raise1pt\hbox{$<$}}}         
\def\gsim{\mathrel{\rlap{\lower2.5pt\hbox{\hskip1pt$\sim$}}
    \raise1pt\hbox{$>$}}}         

\begin{document} 
\title{Nuclear transparency from quasielastic $^{12}$C(e,e'p)}  
\date{\today} 
\author{}\affiliation{}
\author{D.~Rohe}\email[]{Daniela.Rohe@unibas.ch}\affiliation{University of Basel,  CH--4056 Basel, Switzerland} 
\author{O.~Benhar}\affiliation{INFN and Department of Physics, Universit\`a ``La Sapienza", I-00185 Rome, Italy}
\author{C.S.~Armstrong}\affiliation{Thomas Jefferson National Accelerator Facility, Newport News, VA 23606, USA} 
\author{R.~Asaturyan}\affiliation{Yerevan Physics Institute, Yerevan, Armenia}
\author{O.K.~Baker}\affiliation{Hampton University, Hampton, VA 23668, USA}
\author{S.~Bueltmann}\affiliation{University of Virginia, Charlottesville, VA 22903, USA}  
\author{C.~Carasco}\affiliation{University of  Basel,   CH--4056 Basel, Switzerland}
\author{D.~Day}\affiliation{University of Virginia, Charlottesville, VA 22903, USA}  
\author{R.~Ent}\affiliation{Thomas Jefferson National Accelerator Facility, Newport News, VA 23606, USA} 
\author{H.C.~Fenker}\affiliation{Thomas Jefferson National Accelerator Facility, Newport News, VA 23606, USA}
\author{K.~Garrow}\affiliation{Thomas Jefferson National Accelerator Facility, Newport News, VA 23606, USA} 
\author{A.~Gasparian}\affiliation{Hampton University, Hampton, VA 23668, USA}
\author{P.~Gueye}\affiliation{Hampton University, Hampton, VA 23668, USA}
\author{M.~Hauger}\affiliation{University of Basel,   CH--4056 Basel, Switzerland} 
\author{A.~Honegger}\affiliation{University of Basel,   CH--4056 Basel, Switzerland}
\author{J.~Jourdan}\affiliation{University of  Basel,   CH--4056 Basel, Switzerland} 
\author{C.E.~Keppel}\affiliation{Hampton University, Hampton, VA 23668, USA}
\author{G.~Kubon}\affiliation{University of  Basel,   CH--4056 Basel, Switzerland}
\author{R.~Lindgren}\affiliation{University of Virginia, Charlottesville, VA 22903, USA} 
\author{A.~Lung}\affiliation{Thomas Jefferson National Accelerator Facility, Newport News, VA 23606, USA}  
\author{D.J.~Mack}\affiliation{Thomas Jefferson National Accelerator Facility, Newport News, VA 23606, USA} 
\author{J.H.~Mitchell}\affiliation{Thomas Jefferson National Accelerator Facility, Newport News, VA 23606, USA} 
\author{H.~Mkrtchyan}\affiliation{Yerevan Physics Institute, Yerevan, Armenia} 
\author{D.~Mocelj}\affiliation{University of  Basel,   CH--4056 Basel, Switzerland}
\author{K.~Normand}\affiliation{University of  Basel,   CH--4056 Basel, Switzerland}
\author{T.~Petitjean}\affiliation{University of Basel,   CH--4056 Basel, Switzerland} 
\author{O.~Rondon}\affiliation{University of Virginia, Charlottesville, VA 22903, USA} 
\author{E.~Segbefia}\affiliation{Hampton University, Hampton, VA 23668, USA}
\author{I.~Sick}\affiliation{University of Basel,   CH--4056 Basel, Switzerland} 
\author{S.~Stepanyan}\affiliation{Yerevan Physics Institute, Yerevan, Armenia}  
\author{L.~Tang}\affiliation{Hampton University, Hampton, VA 23668, USA}
\author{F.~Tiefenbacher}\affiliation{University of Basel,   CH--4056 Basel, Switzerland} 
\author{W.F.~Vulcan}\affiliation{Thomas Jefferson National Accelerator Facility, Newport News, VA 23606, USA}
\author{G.~Warren}\affiliation{University of Basel,   CH--4056 Basel, Switzerland} 
\author{S.A.~Wood}\affiliation{Thomas Jefferson National Accelerator Facility, Newport News, VA 23606, USA}  
\author{L.~Yuan}\affiliation{Hampton University, Hampton, VA 23668, USA}
\author{M.~Zeier}\affiliation{University of Virginia, Charlottesville, VA 22903, USA}
\author{H.~Zhu}\affiliation{University of Virginia, Charlottesville, VA 22903, USA}
\author{B.~Zihlmann}\affiliation{University of Virginia, Charlottesville, VA 22903, USA}  

\collaboration{E97-006 Collaboration}\noaffiliation

\begin{abstract}
We studied the reaction \ceep\ in quasielastic kinematics at momentum transfers between 0.6 and 1.8 (GeV/c)$^2$ covering the single-particle region. From this the nuclear transparency factors are extracted using two methods. The results are compared to theoretical predictions obtained using a generalization of Glauber theory described in this paper. Furthermore, the momentum distribution in the region of the 1s-state up to momenta of 300~MeV/c is obtained from the data and compared to the Correlated Basis Function theory and the Independent-Particle Shell model.   
\end{abstract} 
\pacs{21.10.-k, 25.30.Dh}
\maketitle 

\section{Introduction} 
Electromagnetic probes are a valuable tool to examine the structure of nuclei. The  distortion of the electron is small and the entire nucleus is probed. In (e,e'p) reactions one can assume in good approximation that only one proton is involved in the primary reaction. However, on the way out of the nucleus the nucleon is subject to reactions which can remove it from the reaction channel under consideration. Such inelastic processes lead to absorption and deflection of the outgoing proton and consequently to a reduction of the detected yield. At high outgoing proton energy, the main process of concern is the reduction of the proton flux. The reduction factor is called nuclear transparency $T_A$. An understanding of the propagation of nucleons in nuclear matter is important for the interpretation of many experiments. 

There are two theoretical approaches to deal with absorption. For low proton kinetic energies ($T_p$ $<$ 200~MeV) an optical potential is used. The parameters of the potential are chosen such as to reproduce the phase shifts and cross sections of  nucleon-nucleus scattering. Its imaginary part accounts for inelastic processes, i.e. absorption. For nucleon momenta larger than 1~GeV/c where  the inelastic part of the free nucleon-nucleon (NN) cross section dominates, Glauber calculations are often used.  They derive the Final State Interaction (FSI) of the propagating nucleon in nuclear medium directly from the elementary nucleon-nucleon cross section $\sigma_{NN}$. In the nuclear medium the N-N cross section has a similar energy dependence as the free one for $T_p$ larger than 1~GeV but is reduced by about 20\%. This value is roughly in agreement with the one obtained using a simplified geometrical model assuming classical attenuation and fitting it to experimental data \cite{Gar02}. At low $T_p$ stronger modifications of the N-N cross section occur due to Pauli Blocking and dispersion effects as it was shown in \cite{steve}. Therefore Glauber calculation are generally thought to have a lower limit of validity in the nucleon kinetic energy. 

In this paper it will be shown that agreement with the data down to $T_p$ = 0.3~GeV can be achieved using an approach based on the correlated Glauber approximation to take into account these effects. A similar approach \cite{steve} was already able to describe the $T_A$ measurement at 180 MeV for $^{12}$C, $^{27}$Al, $^{58}$Ni and $^{181}$Ta within the experimental error bars. Details of the theoretical calculation are given in section \ref{theo}. The theoretical results are compared to data taken as part of the experiment E97-006 and previously measured values for the nuclear transparency described in section \ref{sec_exp}. For the extraction of $T_A$ an improved method was applied in this paper and the transparencies obtained via the standard analysis are given for comparison. The improvement is not only apparent in the better agreement of $T_A$ with the theory but also when comparing the momentum distribution in the region of the 1s$_{1/2}$-state in $^{12}$C with the Independent-Particle Shell Model (IPSM) and the Correlated Basis Function theory (CBF) \cite{Benhar89}.  For this deeper lying state the deviation from the shape of the IPSM momentum distribution is already apparent at momenta $>$ 150~MeV/c.  However, the CBF theory gives a good description up to 250~MeV/c. 

\section{Theory}
\label{theo}

Neglecting many-body contributions to the target electromagnetic current, the
nuclear matrix element entering the definition of the $(e,e^\prime p)$ 
transition amplitude can be written (see, e.g., Ref. \cite{Benhar00})
\beq
M_n({\bf p},{\bf q}) = \langle \Psi^{(-)}_{n {\bf p}} | 
\sum_{\bf k} a_{{\bf k}+{\bf q}}^\dagger a_{\bf k} | \Psi_0 \rangle\ ,
\label{mat:el}
\eeq
where ${\bf p}$ and ${\bf q}$ denote the momentum of the detected proton and the momentum 
transfer, respectively. In the above equation, $a_{\bf k}^\dagger$ and $ a_{\bf k}$ are 
nucleon creation and annihilation operators, $| \Psi_0 \rangle$ is the target ground state,  
satisfying the many-body Schr\"odinger equation 
\beq
H_{\rm A} | \Psi_0 \rangle = E_0 | \Psi_0 \rangle\ ,
\label{schroedinger}
\eeq
and $\Psi^{(-)}_{n {\bf p}}$ is the final scattering state.  

The effect of FSI between the knocked out nucleon (labelled with index 1) and the spectator 
particles can be best analyzed rewriting the A-body nuclear hamiltonian $H_A$ in the 
form
\beq
H_{\rm A} = H_0 + H_{{\rm FSI}} = (H_{{\rm A-1}} + T_1) + H_{{\rm FSI}} \ ,
\label{decomp}
\eeq
where $H_{{\rm A-1}}$ and $T_1$ are the hamiltonian of the recoiling (A--1)-nucleon
system and the kinetic energy operator associated with the struck nucleon,
respectively, while $H_{{\rm FSI}}$ describes the interactions between the
struck nucleon and the spectators. Obviously, PWIA amounts to setting $H_{{\rm FSI}}=0$.

The decomposition of Eq.(\ref{decomp}) can be used to write 
$| \Psi^{(-)}_{n {\bf p}} \rangle$ in the form \cite{Goldberger64}
\beq
| \Psi^{(-)}_{n {\bf p}} \rangle = \Omega^{(-)}_{{\bf p}}
  | \Phi_{n {\bf p}} \rangle \ ,
\label{fin:wf}
\eeq
where the asymptotic state $| \Phi_{n {\bf p}} \rangle$, describing the system in absence 
of FSI, is an eigenstate of $H_0$ of the product form 
\beq
| \Phi_{n {\bf p}} \rangle = | {\bf p} \rangle \otimes
| \varphi_n \rangle \ ,
\label{asy}
\eeq
$| {\bf p} \rangle$ and $| \varphi_n \rangle$ being eigenstates of $T_1$ and 
$H_{{\rm A-1}}$, respectively.

In coordinate space $| \Phi_{n {\bf p}} \rangle$ can be written
(in order to simplify the notation spin indices will be omitted)
\beq
\langle R | \Phi_{n {\bf p}} \rangle =  \Phi_{n {\bf p}} (R) = \sqrt{\frac{1}{V}}\
e^{i{\bf p} \cdot {\bf r}_1} \varphi_n({\widetilde R})\ ,
\label{asy:wf}
\eeq
where $R~\equiv~\{ {\bf r}_1, {\bf r}_2 \ldots ,{\bf r}_A \}$
and ${\widetilde R}~\equiv~\{ {\bf r}_2, \ldots ,{\bf r}_A \}$ specify
the configurations of the full A-particle system and the (A--1)-particle
spectator system, respectively, whereas $V$ denotes the normalization volume. 

Setting $\Omega^{(-)}_{{\bf p}}=1$, which amounts to disregarding the
effects of FSI, and substituting the resulting final state 
into Eq.(\ref{mat:el}), one obtains the
PWIA transition amplitude, that depends upon the initial momentum 
${\bf k} = {\bf p}-{\bf q}$ only.

The scattering operator $\Omega^{(-)}_{{\bf p}}$ in Eq.\ref{fin:wf}
describes the distortion of the asymptotic wave function resulting from rescattering of the
knocked-out nucleon. It can be formally written as \cite{Goldberger64}
\be
\nonumber
\Omega^{(-)}_{{\bf p}} & = & \lim_{t \rightarrow \infty} e^{iH_{\rm A}t} e^{-iH_0t}  \\
& = & \lim_{t \rightarrow \infty} {\widehat T}\
{\rm e}^{-i \int_0^t dt^\prime\ H_{{\rm FSI}}(t^\prime)}\ ,
\label{omega:def}
\ee
where ${\widehat T}$ is the time ordering operator and
\beq
H_{{\rm FSI}}(t) ={\rm e}^{iH_0t}H_{{\rm FSI}}{\rm e}^{-iH_0t}\ .
\eeq

The calculation of $\Omega^{(-)}_{{\bf p}}$ from Eq.(\ref{omega:def}) using 
a realistic nuclear hamiltonian involves prohibitive difficulties. However, when
the kinetic energy carried by the knocked-out proton is large, the
structure of $\Omega^{(-)}_{{\bf p}}$ can be greatly simplified using
 a generalization of the approximation scheme originally developed by Glauber
to describe proton-nucleus scattering \cite{Glauber59}.

The basic assumptions underlying this scheme, generally referred to as correlated
Glauber approximation (CGA) (see, e.g., Ref.~\cite{Ben94} and references therein), are that
i) the fast struck nucleon moves along a straight trajectory, being
undeflected by rescattering processes ({\it eikonal approximation}) and
ii) the spectator system can be approximated by a collection of fixed scattering
centers ({\it frozen approximation}).

Within CGA, $\Omega^{(-)}_{{\bf p}}$ can be written in coordinate space as
\begin{eqnarray}
\nonumber
\langle R | \Omega^{(-)}_{{\bf p}} | R \rangle & = & \Omega^{(-)}_{{\bf p}}(R) 
 = P_z\ \left[ 1 
- \sum_{j=2}^A \Gamma_{p}(1,j) \right. \\
& + & \left. \sum_{k>j=2}^A \Gamma_{p}(1,j)
\Gamma_{p}(1,k) - \ldots \right] \ ,
\label{omega:CGA}
\end{eqnarray}
where the positive $z$-axis is chosen along the eikonal trajectory and the
$z$-ordering operator $P_z$ prevents the occurrence of backward
scattering of the fast struck proton.

The profile function $\Gamma_{p}$, appearing in Eq.(\ref{omega:CGA}) is defined as 
\beq
\Gamma_{p}(1,j) = \theta(z_j - z_1) \gamma_{p}(|{\bf b}_1-{\bf b}_j|)\ ,
\eeq
where the step function preserves causality and $|{\bf b}_1-{\bf b}_j|$ is 
the projection of the interparticle distance on the impact parameter 
plane (i.e. the $xy$ plane). The function $\gamma_{p}(b)$, 
containing all the information on the dynamics of the 
scattering process, is simply related to the NN scattering amplitude 
at incident momentum $p$ and perpendicular momentum transfer $k_t$, $f_{p}(k_t)$, through
\beq
\gamma_{p}(b) = - \frac{i}{2} \int \frac{d^2 k_t}{(2\pi)^2}\
   {\rm e}^{i {\bf k}_t \cdot {\bf b} }\ f_{p}(k_t)\ .
\label{little:gamma}
\eeq

At large $p$, the scattering amplitude $f_{p}(k_t)$ extracted from the measured 
NN cross section is usually parametrized in the
form \cite{nndata}
\beq
f_{p}(k_t) = i\ \sigma_{pN}(1 - i\alpha_{pN})
{\rm e}^{-\frac{1}{2}\frac{k_t^2}{B}}\ ,
\label{para:ampl}
\eeq
where $\sigma_{pN}$ and $\alpha_{pN}$ denote the total cross section and
the ratio between the real and the imaginary part of the amplitude, respectively, 
while the slope parameter $B$ is related to the range of the interaction. In 
case of zero-range, corresponding to $1/B=0$, the impact parameter dependence 
of $\gamma_{p}(b)$ reduces to a two-dimensional $\delta$-function.

The calculation of the scattering operator of Eq. (9) requires the knowledge
of the NN scattering amplitude in the nuclear medium. Pandharipande and Pieper \cite{steve}
have shown that the total NN cross sections in nuclear matter and in vacuum can be
easily related to one another. Their approach, in which the velocity dependence of the
nuclear mean field and Pauli blocking of the final states available to the spectator
nucleons are both taken into account, allows one to obtain the effective total cross
section, ${\widetilde \sigma}_{pN} < \sigma_{pN}$, from the measured differential cross
section $d\sigma_{pN}/d\Omega$. The results of Ref. \cite{steve} suggest
that medium effect reduce the total cross section by as much as $\sim$ 50 \%
and $\sim$ 18 \% at incident energies of .2 and 1. GeV, respectively.

Theoretical studies of the $(e,e^\prime p)$ cross section 
carried out using CGA \cite{Benhar04} show that at large proton momentum the dominant 
FSI effect is a quenching that can be accurately 
described introducing a transparency factor $T_A$, 
written in terms of the scattering operator of Eq.(\ref{omega:CGA}) 
as \cite{Benhar00}
\beq
T_{\rm A} = \frac{1}{Z}\ \int d^3 r\ \rho_p({\bf r})\ 
|{\overline \Omega}^{(-)}_{{\bf p}}({\bf r})|^2 \ .
\label{def:transp}
\eeq
In the above equation $\rho_p({\bf r})$ is the proton density of the target, normalized 
to $Z$, and
\beq
\rho_p({\bf r}_1) |{\overline \Omega}^{(-)}_{{\bf p}}({\bf r_1})|^2 
 = \int d{\widetilde R}\ |\Psi_0(R)|^2\ |\Omega^{(-)}_{{\bf p}}(R)|^2\ .
\label{average:op}
\eeq

The full calculation of the right hand side of Eq.(\ref{average:op}), involves 
3(A--1)-dimensional integrations. It has been carried out for $^4$He and $^{16}$O 
using Monte Carlo techniques and realistic many-body wave functions \cite{Benhar00,Benhar04}. 

Approximations to $T_{\rm A}$ can be obtained expanding the integrand of 
Eq.(\ref{def:transp}) according to 
\be
\nonumber
& & \rho_p({\bf r}_1) |{\overline \Omega}^{(-)}_{{\bf p}}({\bf r}_1)|^2  = 
\ \ \ \ \ \ \ \ \ \ \ \ \ \ \ \ \ \ \ \ \ \ \ \ \ \ \ \ \ \ \ \ \ \ \ \ \ \\
\nonumber
&  &  \ \ \ \ \ \ \ \ \ 1 - \frac{1}{\rho_p({\bf r}_1)} \left[ \ 
 \int d^3r_2 \Gamma(1,2)\rho^{(2)}_{pN}({\bf r}_1,{\bf r}_2) \right.\\
\nonumber
&  &  \ \ \ \ \ \ \ \ \ - \int d^3r_2 d^3r_3 \Gamma(1,2)\Gamma(1,3)
\rho^{(3)}_{pNN}({\bf r}_1,{\bf r}_2,{\bf r}_3)  \\
&  &  \ \ \ \ \ \ \ \ \ + \left. \int d^3r_2 d^3r_3 d^3r_4\  \ldots \ \right] \ ,
\label{expansion}
\ee
where the distribution functions $\rho^{(n)}_{pN \ldots N}({\bf r}_1,{\bf r}_2 \ldots {\bf r}_n)$ 
yield the joint probability of finding the struck proton at ${\bf r}_1$ and the $n-1$ spectator 
nucleons at ${\bf r}_2\ldots{\bf r}_n$. 

The quantity in square brackets in Eq.(\ref{expansion}) describes FSI effects, that lead to a 
departure from the PWIA result $T_{\rm A} = 1$. The contribution of processes 
in which the struck proton undergoes $n-1$ rescatterings involves the $n$-nucleon
distribution. For example, the single rescattering term involves the two-nucleon distribution 
that can be written in the form \cite{corrfcn}
\beq
\rho^{(2)}_{pN}({\bf r}_1,{\bf r}_2) = \rho_p({\bf r}_1)\rho_N({\bf r}_2) 
g({\bf r}_1,{\bf r}_2) \ .
\eeq
The function $g({\bf r}_1,{\bf r}_2)$ describes the effects of dynamical correlations.
Due to the strongly repulsive nature of NN interactions at short range, 
$g(r) \ll 1$ at $r = |{\bf r}_1-{\bf r}_2| \lsim 1$ fm, while 
at large $r$ $g(r) \rightarrow  1$ and the distribution function reduces to the
prediction of the independent particle model.

Equation (\ref{expansion}) shows that inclusion of NN correlations produces a nontrivial
pattern of effects. For example, the presence of $g({\bf r}_1,{\bf r}_2)$ in the two-body
(single rescattering) term enhances the transparency, since the short range repulsion
between the struck particle and the spectator reduces the rescattering probability.
On the other hand, the repulsion between two spectators in the three-body (double 
rescattering) term leads to the opposite effect. 

As short range correlation are known to 
be largely unaffected by shell and surface effects, $g({\bf r}_1,{\bf r}_2)$ can be 
obtained using the Local Density Approximation (LDA), i.e. setting
\beq
g({\bf r}_1,{\bf r}_2) \approx g_{NM}\left[ |{\bf r}_1-{\bf r}_2| \ , \ 
\rho_N\left( \frac{{\bf r}_1+{\bf r}_2}{2} \right) \right] \ ,
\eeq
where $g_{NM}(r,\rho)$ is the radial distribution function calculated in uniform
nuclear matter at constant density $\rho$. 

A similar procedure can be employed to construct distribution functions involving three or more 
nucleons, needed to evaluate the contributions of processes involving more than one 
rescattering. The LDA scheme, combining nuclear matter results and the measured one-body density,
provides a consistent framework to evaluate $T_A$ from Eqs. (\ref{def:transp}) and (\ref{expansion}) provided the
NN scattering amplitude is known.

The theoretical curves shown in Fig. 4 have been obtained using LDA and an effective
NN scattering amplitude written as in Eq. (\ref{para:ampl}). Medium effects have been
taken into account replacing the total cross sections in vacuum with those of
Ref. \cite{steve}, while for the slope parameters the free space values resulting
from the fit of Ref. \cite{oneill}, based on the data of Ref. \cite{NNdata}, have been used. The
third parameter appearing in Eq. (\ref{para:ampl}), namely $\alpha_{pN}$, does not affect the 
transparency factor.

The calculations have been carried out including two-, three- and four-body terms in 
Eq.(\ref{expansion}), corresponding to single, double and triple rescattering. The 
inclusion of three- and four-body contributions to $T_A$ produces a change of
$\sim$ 10\% and $\sim$ 3\%, respectively, in $^{12}$C and $\sim$~15\% and $\sim$~5\%
in $^{197}$Au. 


\section{Experiment and Data Analysis} \label{sec_exp}
The experiment was performed at the Thomas Jefferson National Accelerator Facility (TJNAF) in hall C. Electrons were detected in the High Momentum Spectrometer HMS, protons in the Short Orbit Spectrometer SOS. Data were taken in five kinematics shown in Tab. \ref{kins} on a $^{12}$C target of 2.5\% radiation length thickness. The target thickness was determined by measuring the weight and the size of the target. Its uncertainty is estimated to be 0.3\%. The data cover a range of  Q$^2$ corresponding to $T_p$ = 0.3 to 1.0 GeV.  In the same kinematics data with a liquid hydrogen target were taken. This target consists of an upright standing aluminum cylinder with a diameter of 4 cm and 0.13~mm thick windows. The background contribution from the windows was subtracted using two aluminum targets spaced by the dimension of the cryogenic target. These data serve as a check of the analysis and for the determination of the kinematical offsets to the values given in Tab.~\ref{kins}. The offset to the beam energy was determined to --0.1\% from an independent analysis of \cite{Christy} using a large set of \heep\ data. From the remeasurement of the magnetic dipole field in the magnets located in the arc of the beam line to hall C which are used for the determination of the beam energy, a correction of --0.2\% was found \cite{Mack}. Both offsets were used in the following analysis and contribute to the systematic error. 

\begin{table}[t]
\caption{\label{kins} For the five settings in quasielastic kinematics the averaged momentum transfer $Q^2$, the beam energy $E_e$, the central momentum of the HMS $p_{e'}$ (electron) and of the SOS $p_{p}$ (proton) as well as the central angles $\theta_e$ and $\theta_p$ for HMS and SOS are given.}
\begin{center}
\begin{ruledtabular}
\begin{tabular}{lccccc}
$Q^2$ & $E_e$     &  $p_{e'}$  &  $\theta_e$& $p_{p}$ & $\theta_p$\\
(GeV/c)$^2$ & (GeV) & (GeV/c)  & degree & (GeV/c) & degree \\
\hline
0.59 & 3.298       & 2.95	& 14.4	& 0.85 & 60.3 \\
0.80 & 3.298       & 2.75	& 17.0	& 1.00 & 56.2 \\
1.13 & 3.123       & 2.50	& 22.2  & 1.25 & 49.7 \\
1.53 & 3.298       & 2.40	& 25.4  & 1.50 & 44.6 \\
1.85 & 3.298       & 2.28       & 29.0  & 1.70 & 40.7  \\
\end{tabular}
\end{ruledtabular}
\end{center}
\end{table}

The \heep\ data were also used to determine the fraction of protons absorbed in the target and on their way to the detector system. For this the rates of the reaction \heep\ and H(e,e') in the HMS were compared in a kinematical region covered by both reactions. The result of (0.953 $\pm$ 0.011) agrees with the one calculated from the mean free path and the material thicknesses.

The analysis of the \heep\ data revealed that for momenta larger than 1~GeV/c the protons punch through the collimator which consists of 2.5 inch tungsten. This leads to additional energy loss of 50 to 300 MeV for less than 3\% of the events. The effect was included in the simulation (see below).

The \heep\ data are in good agreement with previous results. The momentum acceptance relative to the central momentum $\Delta p/p$ was --8.8\% to 10.4\% for the HMS and $\pm$ 15\% for the SOS. It was determined using overlapping spectra from inclusive electron scattering on $^{12}$C in the inelastic regime where the cross section is a smooth function of the momentum. Both spectrometers were fixed at 20$^{\circ}$ and the central momentum was varied in steps of 0.1~GeV/c. This guarantees an overlap between the settings. A correction of the order of $\pm$ 2\% was applied to the yield as a function of $\Delta p/p$. 

The spectra from the \heep\ data were compared to the Monte Carlo simulation SIMC of the hall C collaboration and good agreement were found. The simulation describes the electron beam entering the target (including the rastering of the beam over the target used to avoid local heating) and follows the particles from the reaction vertex to the detector system. It employs transfer matrices for the spectrometers and takes energy loss and multiple scattering in material into account. Radiative corrections are calculated according to \cite{Ent01}. The effect of Coulomb distortion on the electron wave function in the vicinity of heavy nuclei is small at the momenta considered here ($\approx$ 1\%). It was taken into account by using the effective momentum approximation. The angular and momentum resolutions were adjusted to reproduce best the measured spectra. For this the spectra of  the missing energy \emiss\ and missing momentum \pmiss\ are suitable in particular. A resolution (FWHM) in \emiss\ of $\approx$ 7 MeV and in \pmiss\ of $\approx$ 9 MeV/c is achieved.  From the measured kinematic variables, the missing energy is reconstructed
according to
\begin{equation} \label{Em}
E_m = E_e - E_e' - T_p - T_R .
\end{equation}
Here, $E_e$ is the beam energy and $E_{e'}$ and $T_p$ are the (kinetic) energies of the outgoing electron and proton, respectively. $T_R$ is the kinetic energy of the (undetected) recoiling (A--1)-system, which is reconstructed from \pmiss\ using the spectator model. The missing energy can be identified with the removal energy $E$ of the nucleon. In the single-particle region it is the energy needed to remove the nucleon from a particular state within the nucleus. In Plane Wave Impuls Approximation (PWIA) the initial momentum $k$ of the nucleon bound in the nucleus is equal to the opposite of the missing momentum ${\bf p}_m$ = ${\bf q} - {\bf p}$ where ${\bf q}$ is the momentum transfer from the electron to the nucleon. In PWIA the 6-fold differential cross section 
\begin{equation} \label{cross}
\frac{d^6\sigma}{dE\,dE'\,d\Omega_{e}\,dE_p\,d\Omega_{p}} = K\,\sigma_{ep} S(E,{\bf k})\,T_A(Q^2) 
\end{equation}
factorizes into an elementary e--p cross section $\sigma_{ep}$ and the spectral function $S(E,{\bf k})$ containing the information about the nuclear structure. The spectral function is the probability to find in the nucleus a nucleon with energy $E$ and momentum $k$. 

The choice of cross section $\sigma_{ep}$ for the scattering of an electron on a bound nucleon is not unique. Since the off-shell form factors are not known the same hadronic currents as in the case of free nucleons are used but do not lead to the same result. The deviation between different choices increases with $E$ and $k$. For the result of the present work, which is restricted to the single-particle region, the ambiguity is small. To be consistent with previous analyses the off-shell prescription $\sigma_{ep}^{cc1}$ of ref. \cite{Forest83} was used. The deviation to $\sigma_{ep}^{cc2}$ \cite{Forest83} and to  $\sigma_{ep}^{cc}$ \cite{Rohe04} is less than 1.5\% for the five settings in the \emiss-\pmiss\ region used for this analysis. 

As parameterization for the electromagnetic form factors of the proton we use the dispersion-theoretical analysis based on vector meson dominance model of \cite{Mergell96}. The change on the cross section at the highest $Q^2$ of the present work is less than 0.5~\% when the result from the double-polarization method for the $G_{ep}/G_{mp}$ ratio \cite{Jones00,Gayou02} is employed.

\begin{figure}[t] 
\begin{center}
\includegraphics[width=7.5cm,clip]{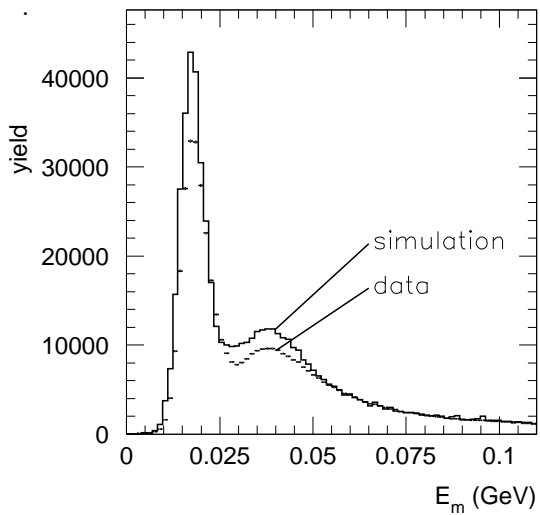}
\end{center}
\caption{\label{qu_Em}Comparison of the experimental yield obtained as a function of \emiss\ in the single--particle region to the Monte Carlo simulation (solid) for the kinematics at $Q^2$ = 1.13 (GeV/c)$^2$ (s. Tab. \ref{kins}). Both contributions contain radiative processes.}
\end{figure}

In the framework of PWIA the nuclear transparency enters as correction factor on the right hand side of Eq.(\ref{cross}). Experimentally it is obtained by comparing the measured yield $N^{exp}(E_m,{\bf p}_m)$ to the simulation $N^{sim}(E_m,{\bf p}_m)$ integrated over the same phase space $V$
\begin{equation} \label{extr_trans}
T_A(Q^2) =  \frac{\int_V d{\bf p}_m\, dE_m\, N^{exp}(E_m,{\bf p}_m)}{\int_V d{\bf p}_m\, dE_m\, N^{sim}(E_m,{\bf p}_m)}.
\end{equation} 
To restrict the analysis to the single-particle region cuts in \emiss\ $\leq$ 0.08~GeV and \pmiss\ $\leq$ 0.3~GeV/c were applied. The same cuts were also used in most of the previous experiments. 

In the simulation a spectral function has to be chosen which correctly describes the experimental distribution well in the region of interest. In Fig. \ref{qu_Em} the \emiss-spectrum of the experiment (dots) is compared to the simulation using an IPSM spectral function as input (solid curve). This spectral function factorizes into an $E$ and $k$ distribution for each orbit. It was also used for the analyses of the experiment NE-18 \cite{Nei95} at SLAC and experiments at TJNAF \cite{Abb98,Gar02}. The $E$-distribution is described by a Lorentzian and the momentum distribution is derived from a Wood-Saxon potential whose parameters are adjusted to data measured at Saclay \cite{Mou76}. For this figure $T_A$ = 0.6 in the simulation was used. The difference in the yield reflects the depletion of the single-particle region due to short-range correlations (SRC). To account for this the simulated yield for carbon is divided by a factor of $\epsilon^{SRC}$ 1.11 $\pm$ 0.03. This factor was used in all the previous analyses and was estimated from an early examination of N-N correlations in $^{12}$C and $^{16}$O \cite{Sick93,Orden80}. The inverse of $\epsilon^{SRC}$ corresponds to the occupation number, which modern many-body theories predict to be lower, $\approx$ 80\% \cite{Benhar89,Mue95}. This number is larger than the spectroscopic factor giving the occupation of a state; it contains the background term induced by correlations which cannot be attributed to a specific orbit \cite{Ben90,Mue04}. The above considerations emphasize that the number of nucleons missing due to absorption in the nucleus cannot be distinguished from the depletion of the single-particle orbits due to SRC.  

\begin{figure}[tbh] 
\begin{center}
\includegraphics[width=7.5cm,clip]{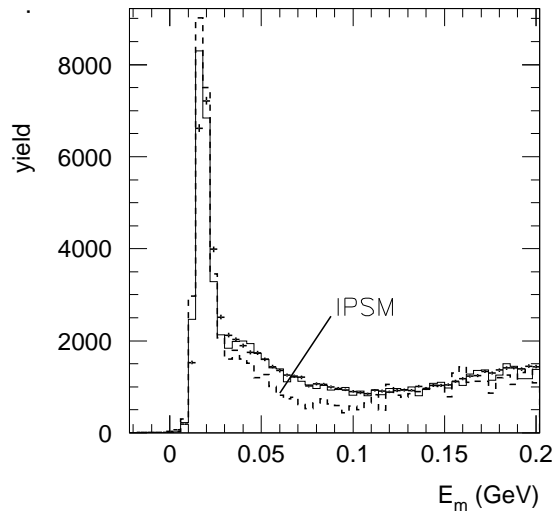}
\caption{\label{qutail} \emiss-distribution for a slice in \pmiss\ of 250 $\pm$ 50~MeV/c in the quasi-elastic kinematics at $Q^2$ = 1.53 (GeV/c)$^2$. The experimental yield  is shown as data points, the other curves are simulated taking radiative processes into account. Using the IPSM model results in the lower curve (dashed). For the distribution (solid) the spectral function of \cite{Ben94} is used as input for the Monte Carlo.}
\end{center}
\end{figure}

To improve upon this approach a spectral function containing SRC from the beginning is employed in Eq.(\ref{extr_trans}). This spectral function is composed of a part due to SRC which accounts for 22\% of the total strength as calculated in LDA, and the IPSM spectral function mentioned above but reduced by a factor (1 -- 0.22) to ensure normalization. This approach circumvents the application of the extra factor $\epsilon^{SRC}$ which is a poor approximation because the amount of strength due to SRC is increasing with \pmiss. Its limitation is demonstrated in Fig. \ref{qutail}. For a \pmiss\ bin of (250 $\pm$ 50) MeV/c the experimental \emiss-distribution is compared to the simulation using the IPSM spectral function (including the correction  $\epsilon^{SRC}$) and the CBF spectral function. It is obvious that the IPSM fails already at moderate \pmiss\ and \emiss\ whereas the CBF theory is in good agreement with the data. Note that the increase of the yield at larger \emiss\ is not due to $S(E,k)$, but is due to bremsstrahlung shifting events from low to high \emiss.     
 
\section{Results} \label{results}
\begin{figure}[t] 
\begin{center}
\includegraphics[width=7.5cm,clip]{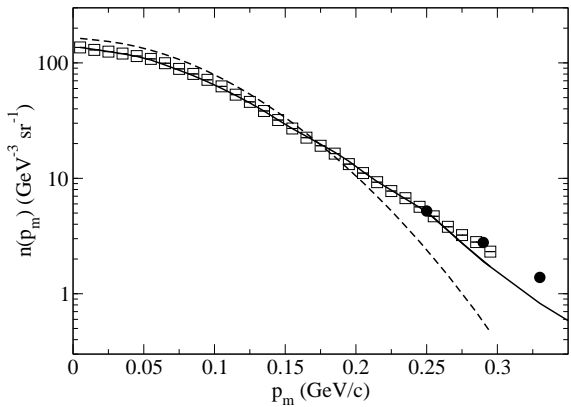}
\caption{\label{s_mom}Momentum distribution in the region of 0.03~GeV $<$ \emiss\ $<$ 0.08~GeV obtained from the data taken in the kinematics of Tab. \ref{kins} (squares), the CBF theory (solid) and the IPSM (dashed). Three data points (circles) are from data focusing on the high \pmiss\ region \cite{Rohe04,Rohe04a}.}
\end{center}
\end{figure}  
When extracting $T_A$ using Eq.(\ref{extr_trans}) good agreement between Monte Carlo simulation and data is required. Spectra of kinematical quantities like momenta and angles depend mainly on the capability to map the optics of the spectrometers. These were verified using the \heep\ data.  The comparison with the simulated \emiss\ and \pmiss\ distribution then reflects the quality of the spectral function used as input for the simulation. The agreement in the \emiss\ distribution can easily be judged from the spectrum given in Fig. \ref{qu_Em}. In contrast to previous works, where the same spectral function was used, the width of the $1p_{3/2}$ state had to be reduced from 5~MeV to 1~MeV. This value is closer to the expectation of a vanishing width for the  $1p_{3/2}$ state. The momentum distribution of the $1s_{1/2}$ state was obtained from the data. For this the \pmiss\ distribution was split into 10~MeV/c bins up to 300~MeV/c. The data then were compared bin  for bin to the simulation using the CBF spectral function. Simultaneously radiative corrections were taken into account by the de-radiation technique. The ratio was taken as a correction factor to the momentum distribution. The total normalization was adjusted to the simulation (using $T_A$ = 0.6) which compensates for small differences in $T_A$ in the five kinematics. The result obtained from the data (squares) is shown in Fig. \ref{s_mom} compared to the momentum distribution from the CBF theory (solid blue line) and from the IPSM (dashed black line) integrated over the same region in \emiss\ (0.03 -- 0.08~GeV). The statistical error bars are smaller than the symbols. The systematic error is given below. Fig. \ref{s_mom} shows clearly that the IPSM fails already above \pmiss\ = 0.15~GeV/c. This is in contrast to the measurements done at NIKHEF on various nuclei (s. refs. \cite{Frullani84,Lapikas93}) where good agreement was found with the momentum distribution derived from a Wood-Saxon potential up to momenta of 250~MeV.  In these experiments only valence states at the Fermi edge were examined. For deeper lying states like the $1s_{1/2}$ state the influence of SRC leads to modifications of the shape already at moderate \pmiss. 

Since no correction factor $\epsilon^{SRC}$ is applied to the IPSM momentum distribution in figure \ref{s_mom} it exceeds the experimental momentum distribution by $\approx$ 20\% at small \pmiss. The deviation at large \pmiss\  is not a too serious problem  because most of the yield used for the extraction of $T_A$ comes from smaller \pmiss. The CBF theory gives a resonable description to the data up to \pmiss\ $\approx$ 0.25~GeV/c which is close to the upper limit used in the $T_A$ analysis. At higher \pmiss\ the CBF theory underestimates the data in this \emiss\ region. 

The data points shown as full circles were also taken in the E97-006 experiment but with the aim to examine SRC at high \emiss\ and \pmiss\ \cite{Rohe04,Rohe04a}. For this a spectral function was extracted from the data. The momentum distribution was obtained by integrating the spectral function over the given \emiss\ region. These data are not subject of this article but the agreement in the overlap region of both data sets provides an important confirmation of consistency using different analysis methods. 

\begin{table}[th]
\caption{\label{transp_res}Nuclear transparency for carbon obtained using the IPSM  and the CBF spectral function in the analysis. The error given is the quadratic sum of statistical and systematic uncertainty neglecting the model--dependent error.}
\begin{center}
\begin{ruledtabular}
\begin{tabular}{lcc}
$T_p$ (GeV)  & $T_A$ (IPSM) & $T_A$ (CBF)\\
\hline
0.328	&     0.598 $\pm$ 0.023 &    0.641 $\pm$ 0.025\\
0.433	&     0.581 $\pm$ 0.023 &    0.628 $\pm$ 0.025\\
0.625	&     0.566 $\pm$ 0.022 &    0.605 $\pm$ 0.024\\
0.830	&     0.561 $\pm$ 0.022 &    0.593 $\pm$ 0.023\\ 
1.00	&     0.553 $\pm$ 0.022 &    0.591 $\pm$ 0.023\\    
\end{tabular}
\end{ruledtabular}
\end{center}
\end{table}

The results for the nuclear transparency $T_A$ obtained using the IPSM and the CBF spectral functions in Eq.(\ref{extr_trans}) are shown in table \ref{transp_res}. When using the CBF theory the results are about 5\% higher than the results using IPSM. This  can be traced back to the larger influence of SRC predicted by the CBF theory, the correlated strength of which is in agreement with the strength found at high $E$ and $k$ by experiment \cite{Rohe04a}. From the $\epsilon^{SRC}$ quoted above, which corresponds to a depletion in the single-particle state of 0.1, and the contribution of SRC in the CBF  theory (0.22) one could naively have expected a larger difference between the two analyses. The deviation is not so large due to strength from SRC which also contributes to the single-particle region.

It should be noted that the results for the momentum distribution and the nuclear transparency factor might depend slightly on the kinematics chosen. In general the yield obtained in perpendicular kinematics exceeds the one measured in parallel kinematics due to additional reaction mechanisms present in perpendicular kinematics \cite{Barbieri04}. In the single-particle region (below the Fermi momentum) a difference of 6~\% is expected \cite{Kelly96,Nikolaev95}.  

In the region of $Q^2$ = 0.15 - 0.6 (GeV/c)$^2$ refs. \cite{Ulmer87} and \cite{Dutta00} found an excess of the transverse response  40~\% in $^{12}$C for \emiss\ $>$ 0.025 GeV compared to a free proton. At higher $Q^2$ the transverse response quickly decreases. For the data point at $Q^2$ = 0.59 GeV/c)$^2$ this would lead to an increase of the transparency factor of $\approx$ 9~\%. To search for this effect in the data we extracted the transparency factor from the 1p-state alone (\emiss\ $<$ 0.025 GeV). We find a decrease of $T_A$ by 4~\%, i.e. a two times smaller effect. At $Q^2$ = 0.8 (GeV/c)$^2$ the effect is 1~\% and decreases futher for higher $Q^2$.

\begin{figure}[t] 
\begin{center}
\includegraphics[width=7.5cm,clip]{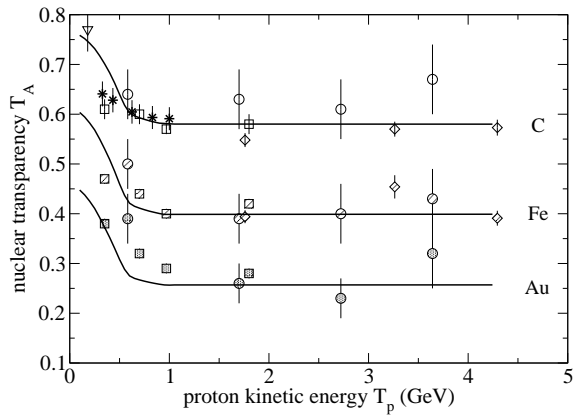}
\caption{\label{res_trans}Nuclear transparency $T_A$ for C, Fe and Au as a function of the proton kinetic energy $T_p$ compared to the correlated Glauber calculations (solid lines). The data indicated by circles are from the NE18--experiment at SLAC \cite{Nei95}, squares and diamonds are Jlab data of \cite{Abb98} and \cite{Gar02} and from Bates \cite{Garino92} (triangle down). The result indicated by stars is obtained with the correlated spectral function of \cite{Ben94}.}
\end{center}
\end{figure}     

The error given in table \ref{transp_res} is the quadratic sum of the statistical and systematic uncertainties where the latter dominates with 3.9\% in total. It consists of the already mentioned uncertainties in the target thickness (0.3\%), the charge measurement (1\%), proton transmission (1.1\%) as well as the uncertainties in several correction factors that have to be applied to the experimental yield like dead time and detector efficiency (2\%). The stability of the result against different cuts in the data was conservatively estimated to 2\%. For the simulation, in particular the phase space, an error of 2\% was taken into account. The influence of the choice of the kinematical offsets was tested with the above mentioned second set of offsets corresponding to a difference in energy of 0.1\%. It leads to differences in the result of less than 0.5\%. For the fluctuations from run to run an uncertainty of 0.5\% was estimated. The systematic error does not contain the model-dependent uncertainty. For the choice of the spectral function an error of 2\% is taken into account which reflects mainly the agreement between data and simulation discussed above. The error in the correction factor $\epsilon^{SRC}$ of 3\% has to be applied only in the case of the analysis using the IPSM spectral function. To estimate the influence of the off-shell cross section the analysis was repeated using the cc2-version of \cite{Forest83} and the cc-version of \cite{Rohe04,Rohe04a}. The latter leads to almost the same result as the cc1-version of \cite{Forest83}. An error of 2\% was taken into account. The contribution of bremsstrahlung to the data amounts to 33\%. The uncertainty in the correction due to internal (external) bremsstrahlung was estimated to 2\% (1\%). Summing these uncertainties quadratically leads to a model-dependent uncertainty of 4.7\% (3.6\%) for the analysis using the IPSM (CBF) spectral function.    

The measured nuclear transparency $T_A$ (solid symbols) for $^{12}$C is shown in Fig. \ref{res_trans} as a function of the kinetic energy of the proton together with previous results obtained at SLAC \cite{Nei95} (circles)  and Jlab \cite{Abb98,Gar02} (squares and diamonds). The error bars shown in the figure contain the statistical and systematic uncertainty but not the model-dependent error. This applies also to the data points of the previous works. Since the previous experiments were analyzed using the same assumption and ingredients the model-dependent error is the same for them, while it is somewhat lower in the case of using the CBF spectral function.

The solid lines drawn in Fig. \ref{res_trans} are the result of the theory presented in this paper. For comparision also results from previous experiments \cite{Gar02,Nei95,Abb98} for iron and gold are shown. For all three nuclei and large proton kinetic energy ($>$ 1.5~GeV) the theory describes the data well within the error bars. At low energy there is remarkable agreement between theory and the experimental results obtained using the CBF spectral function. The two data points at the lowest $T_p$ for $^{12}$C could indicate a deviation from the prediction, but considering the model-dependent error bar no firm conclusion can be drawn. With the standard analysis the experimental results are $\approx$ 5\% too low but in agreement with previous analyses using the same ingredients. On the other hand the data points for gold seem to exceed the theory.  For these analyses a correction factor 1/$\epsilon^{SRC}$ = 0.78 was used \cite{Nei95,Abb98}. If one would have used the CBF spectral function the results would be lowered by $\approx$ 7\% and thus closer to the theory.    

\acknowledgements
This work was supported by the Schweizerische Nationalfonds (SNF), the US Dept. of
Energy  and the US National Science Foundation.

One of the authors (OB) wishes to express his gratitude to S.C. Pieper for a very useful correspondence and for providing tables of the medium modified NN cross sections. Many illuminating discussions with V.R. Pandharipande are also gratefully acknowledged.

\end{document}